# Reducing bias in dual flip angle $T_1$-mapping in human brain at 7T

Hampus Olsson[1], Mads Andersen[2], Jimmy Lätt[3], Ronnie Wirestam[1], Gunther Helms[1]

[1]*Dept. of Medical Radiation Physics, Clinical Sciences Lund, Lund University, Lund, SE.*
[2]*Philips Healthcare, Copenhagen, Denmark.* [3] *Center for Medical Imaging and Physiology, Skane University Hospital, Lund, Sweden.*

**Running head:** Dual flip angle $T_1$-mapping at 7T

**Word count:** 4927

**Contact information:**

Hampus Olsson, M.Sc.
Lund University Clinical Sciences, Dept. of Medical Radiation Physics
Barngatan 4, Skane University Hospital
Lund, SE 221 85
+46737097806
Hampus.Olsson@med.lu.se

## Abstract

**Purpose:** To address the systematic bias in whole-brain dual flip angle (DFA) $T_1$-mapping at 7T by optimizing the flip angle pair and carefully selecting RF pulse shape and duration.

**Theory and Methods:** Spoiled gradient echoes can be used to estimate whole-brain maps of $T_1$. This can be accomplished by using only two acquisitions with different flip angles, i.e., a DFA-based approach. Although DFA-based $T_1$-mapping is seemingly straightforward to implement, it is sensitive to bias caused by incomplete spoiling and incidental magnetization transfer (MT) effects. Further bias is introduced by the increased $B_0$ and $B_1^+$ inhomogeneities at 7T. Experiments were performed to determine the optimal flip angle pair and appropriate RF pulse shape and duration. Obtained $T_1$ estimates were validated using inversion recovery prepared EPI and compared to literature values. A multi-echo readout was used to increase SNR, enabling quantification of $R_2^*$ and susceptibility, $\chi$.

**Results:** Incomplete spoiling was observed above a local flip angle of approximately 20°. An asymmetric gauss-filtered sinc pulse with a constant duration of 700 μs showed a sufficiently flat frequency response profile to avoid incomplete excitation in areas with high $B_0$ offsets. A pulse duration of 700 μs minimized effects from incidental MT.

**Conclusion:** When performing DFA-based $T_1$-mapping one should (i) limit the higher flip angle to avoid incomplete spoiling, (ii) use a RF pulse shape insensitive to $B_0$ inhomogeneities and (iii) apply a constant RF pulse duration, balanced to minimize incidental MT.

**Keywords: $T_1$; 7T; longitudinal relaxation; spoiled gradient echo; dual flip angle; human brain**

# Introduction

The longitudinal relaxation time, T$_1$, is related to the concentration of iron and myelin in the brain(1,2). $T_1$ is thus a relevant parameter in the process of normal aging(3) but also in pathologies such as multiple sclerosis and Parkinson's disease(4-6). Quantification of $T_1$ on a voxel-wise basis allows for reproducible maps of a tissue-specific metric that can be compared longitudinally(7) and across sites(8). The increase in contrast compared to conventional weighted images, due to the removal of confounding contrast sources, may allow for visualization of otherwise invisible pathologies, and is more directly related to tissue properties(9).

A seemingly straightforward and undemanding way to perform quantification of T$_1$ is by varying the flip angle of a gradient echo sequence under spoiled steady-state conditions(10-12). A two-point measurement, i.e., a dual flip angle (DFA) approach, is sufficient to determine T$_1$ as well as the signal amplitude reflecting proton density (PD). The DFA approach is the fastest method to acquire high resolution whole-brain maps of T$_1$ and has been applied extensively on clinical MR systems and dedicated processing tools, like the hMRI toolbox(13) have been made available. Besides T$_1$, this toolbox can also provide maps of $R_2^*$, proton density (PD), and magnetization transfer (MT) saturation.

Studies employing DFA-based T$_1$-mapping of the whole-brain at 7T have, to our knowledge, not previously been published although one study has employed three flip angles to map T$_1$ in a single 2D slice (14). When implementing a DFA protocol at 7T, one must consider the reduced homogeneity of primarily the transmitted radio frequency (RF) field, $B_1^+$, but also $B_0$, as well as the prolonged T$_1$ compared to 3T. The DFA approach for quantification of T$_1$ is very sensitive to $B_1^+$ inhomogeneities, thus accurate flip angle mapping becomes more important at 7T for performing a post-hoc correction of T$_1$ estimates(15). The prolongation of T$_1$ entails an increase in TR and to avoid dead time, a multi-echo readout can be accommodated within TR. This is a time-efficient approach to increase SNR through averaging across echoes(16) while at the same time enabling quantification of $R_2^*$ and magnetic susceptibility, χ.

In this study, the procedure to optimize a DFA protocol for whole-brain T$_1$-mapping at 7T is described. Primarily, the choice of flip angles as well as RF pulse shape and duration are discussed and motivated by minimizing noise propagation and addressing systematic biases in the $T_1$ estimation. The investigated sources of bias were residual transverse coherences, reduced flip angles due to $B_0$ inhomogeneities, as well as incidental on-resonance magnetization transfer (MT) effects on the longitudinal magnetization.

The absorption of RF by the bound macromolecular pool depends on $(B_1^+)^2$ and will thus cause partial saturation of the bound macromolecular pool(17). When the bound pool is more saturated than the bulk water, the observed steady state is lowered by MT to the bound pool, which in turn is interpreted as an increase in $T_1$. Such MT induced changes of $T_1$ have been reported for 3T(18), but can be addressed by choice of pulse shape and duration.

A finalized DFA $T_1$-mapping protocol with multi-echo readout is produced, from which $R_2^*$, χ, and PD estimates are also derived by the multi-parametric hMRI toolbox(13). $T_1$ values, estimated using a rational approximation of the Ernst equation(15), are reported in subregions of normal brain tissue for nine healthy volunteers and compared to literature values. Validation experiments were performed, both in a multi-$T_1$ phantom and in-vivo, using an inversion recovery (IR) prepared EPI sequence. The total measuring time for a whole-brain multi-parameter map with isotropic 0.9 mm resolution was under 8 min including flip angle mapping.

**Theory**

**$T_1$ estimation via a rational approximation of the Ernst equation**

The dependence of the steady-state signal on the flip angle, α, in a spoiled gradient echo with perfectly spoiled transverse coherences is given by the Ernst equation:

$$S(\alpha) = |A| \sin(\alpha) \frac{1 - \exp(-\mathrm{TR}/T_1)}{1 - \cos(\alpha) \exp(-\mathrm{TR}/T_1)} \qquad [1]$$

where $A$ denotes the complex signal amplitude at TE, that is, the signal obtained for $\alpha = \pi/2$ and fully relaxed conditions, i.e. TR $\gg$ $T_1$. $A$ is a function of TE, thus allowing quantification of $R_2^*$ and $\chi$. For small flip angles ($\alpha \ll 1$) and short TR $\ll T_1$, the expression in Eq. [1] can be approximated by(15,19):

$$S(\alpha) \approx |A|\alpha \frac{TR/T_1}{\alpha^2/2 + TR/T_1}. \qquad [2]$$

This equation immediately shows the change from $T_1$-weighting ($T_1$-w) to PD-w with decreasing flip angles when $\alpha^2/2$ becomes smaller than $TR/T_1$ and the fraction of the right-hand term approaches unity.

The Ernst angle, $\alpha_E$, defined by $\cos \alpha_E = \exp(-TR/T_1)$, for a given TR and $T_1$, is correspondingly approximated by

$$\alpha_E \approx \sqrt{2TR/T_1}. \qquad [3]$$

Lastly, Eq. [2] can be rewritten as

$$\frac{S(\alpha)}{\alpha} = |A| - \frac{T_1}{2TR} \cdot S(\alpha)\alpha \qquad [4]$$

revealing a linear relationship when $S(\alpha)/\alpha$ is plotted against $S(\alpha) \cdot \alpha$. $|A|$ can thus be derived as the intercept and $T_1$ from the slope of a regression line. This also provides a simple way to visually inspect variable flip angle (VFA) data for systematic biases such as those resulting from residual transverse coherences(20).

In the DFA-based $T_1$ estimation, two flip angles ($\alpha_{T1}$, $\alpha_{PD}$) are used to provide two signals $S_{T1}$ and $S_{PD}$, which are predominantly $T_1$-w and PD-w. Use of the nominal flip angles in Eq. [2] ($\alpha_{nom}$, as defined on the user interface) yields an apparent $T_1$(15),

$$T_{1,app} = 2TR \frac{S_{PD}/\alpha_{PD} - S_{T1}/\alpha_{T1}}{S_{T1}\alpha_{T1} - S_{PD}\alpha_{PD}}, \qquad [5]$$

as well as an apparent signal amplitude,

$$|A_{app}| = \frac{S_{T1}S_{PD}(\alpha_{T1}/\alpha_{PD} - \alpha_{PD}/\alpha_{T1})}{(S_{T1}\alpha_{T1} - S_{PD}\alpha_{PD})}. \qquad [6]$$

Note that apparent stands for "without correcting for $B_1^+$ inhomogeneities". Correspondingly, replacing $T_1$ with $T_{1,app}$ in Eq. [3] yields the *apparent* $\alpha_E$, that is, the nominal flip angle at which the signal maximum would be observed in this voxel.

The propagation of noise from the signals $S_{T1}$ and $S_{PD}$ into the calculated $T_{1,app}$ is minimized for the following flip angles(19):

$$\alpha_{T1,opt} \approx \alpha_E \cdot 2.414, \qquad [7]$$

$$\alpha_{PD,opt} \approx \alpha_E / 2.414. \qquad [8]$$

**Correcting for $B_1^+$ inhomogeneities**

In the presence of $B_1^+$ inhomogeneities, the local flip angle is described by the transmit field bias, $f_T$, as

$$\alpha = f_T \cdot \alpha_{nom.} \qquad [9]$$

Inserting this into Eq. [4], $f_T$ appears in the intercept and the slope, yielding the $f_T$-corrected parameter values (15,19):

$$T_1 = T_{1,app}/f_T^2 \qquad [10]$$

$$|A| = |A_{app}|/f_T. \qquad [11]$$

Thus, $f_T$ cannot be derived from a VFA at low flip angles but must be mapped independently. From $|A|$, PD can be derived by numerically approximating the receive field(13).

**Methods**

Experiments were performed on an actively shielded 7T MR system (Achieva, Philips Healthcare, Best, NL), using a head coil with two transmit and 32 receive channels (Nova Medical, Wilmington, MA). Healthy adult subjects were scanned after giving informed written consent as approved by the regional Ethical Review Board.

A non-selective 3D multi-echo spoiled gradient echo ("T1-FFE") sequence was used. Sagittal volumes of 230×230 mm$^2$ in-plane FOV and 200 mm in the right-left direction (with some variation due to subject size) were acquired at isotropic voxel size of (0.9 mm)$^3$ with readout in the head-feet direction. Within TR = 18 ms, eight equidistant gradient echoes were acquired with fat and water in-phase (at multiples of TE = 1.97 ms) at a bandwidth of 670 Hz/px using alternating readout gradient polarity to reduce eddy currents, peripheral nerve stimulation, and gradient heating. The scan time was 3:23 min using a SENSE-factor(21) of 2 in both phase-encoding directions and an elliptical k-space coverage.

A series of experiments were performed to determine the optimal settings for the PD-w and T$_1$-w acquisitions. A variable flip angle experiment was performed to identify bias caused by incomplete spoiling(20) and to minimize noise propagation(19). To mitigate effects of B$_0$ inhomogeneities, Bloch equation simulations were performed to determine a suitable RF pulse shape with a flat frequency response profile for the excitation. The duration of the RF pulse was varied to reduce the effect of incidental on-resonance MT effects on the T$_1$ maps(18).

The dual refocusing echo acquisition mode (DREAM)(22) optimized for 7T(23) was used for B$_1^+$ mapping. Eighty transverse slices with FOV of 240×240 mm$^2$, voxel size of 3.75×3.75×3.50 mm$^3$, readout bandwidth of 4796 Hz/px, were acquired in 10 s per scan. Three B$_1^+$ maps with STEAM preparation angle of 25°, 40° and 60° were acquired to account for the variation of $B_1^+$ at 7T(24). In brief, voxels showing low SNR (at local α < 25°) or bias (at local α > 50°) were masked out and the remaining overlapping maps averaged, resulting in a single composite $f_T$ map(24).

**Data processing**

DICOM images were exported, pseudo-anonymized and converted to 4D NIfTI files using an in-house modification of the dicm2nii tool(25). In brief, voxel intensities were scaled back to physical signal [a.u.] using private DICOM tags, phase maps were converted to radians, and spatial dimensions were re-ordered to conform to radiological convention (right-left) in the standard transverse orientation.

Processing was performed in FSL(26) and MATLAB. For each scan, voxel intensities were averaged across TEs. Rigid-body co-registration of the separate scans was performed using

FLIRT(27,28). The brain mask was derived from the PD-w scan using BET(29). $T_1$ maps were calculated using Eq. [5] and corrected post-hoc using the composite DREAM $f_T$ map.

To account for any residual effects of imperfect spoiling on the $T_1$ estimates arising in high $B_1^+$ areas, the $f_T$ map was modified based on simulations performed by Baudrexel et al.(30). In this work, voxel-wise correction factors for incomplete spoiling are calculated based on the RF phase increment which was 150° in this work.

For PD and $R_2^*$ maps, processing was done by the hMRI toolbox to exploit the built-in functions of data-driven receive field estimation and normalization of PD maps to white matter (WM) where the mean PD value is assumed to be 69 percentile units (p.u.)(8) as well as the use of ESTATICS(31) for $R_2^*$ calculations. The Multi-Scale Dipole Inversion (MSDI) algorithm (32) was used for QSM on the $T_1$-w multi-echo dataset.

**Experiment 1: Nominal flip angle pair and spoiling bias**

The nominal flip angle was varied from 4° to 32° in increments of 4° with otherwise constant parameter settings. This VFA experiment allowed to visualize the signal bias caused by incomplete spoiling through deviations from linearity in Eq. [4] and minimize noise propagation as in Eqs. [7]-[8].

Incomplete RF spoiling depends on the RF phase increment and can affect the signal in a spoiled steady-state gradient echo acquisition and thus lead to systematic bias in the $T_1$ estimation(33). The effect increases with the local flip angle and is negligible beneath a certain threshold. To identify the onset of bias due to incomplete spoiling, ROI analysis was performed in a high and low $B_1^+$ area, respectively. This way, a suitable upper limit on $\alpha_{T1}$ could be determined.

A voxel-wise whole-brain map of the apparent Ernst angle was calculated from the slope of the linear regression, excluding nominal flip angles higher than the upper limit set on $\alpha_{T1}$. To achieve a compromise regarding noise progression over the whole brain, choice of $\alpha_{T1}$ and $\alpha_{PD}$ was based on the whole-brain median of the apparent Ernst angle.

Recognizing that some incomplete spoiling likely occurred in high $B_1^+$ areas regardless of the conservative choice of flip angles, further steps were taken to reduce bias in the $T_1$ estimation. The correction factor derived by Baudrexel et al.(30) was superimposed onto the composite $f_T$ map. This modified $f_T$ map was then used for correction of $T_1$ estimates.

**Experiment 2: RF pulse shape**

To avoid a decrease in the local flip angle in areas with a high proton resonance frequency ($f_0$) offset, different RF pulse shapes, available at the MR system, were evaluated. High $B_1^+$ amplitudes (close to 20 µT) are to be avoided to preclude incidental MT effects (experiment 3), which comes at the cost of reduced bandwidth of the frequency response. Thus, the flatness of the frequency response profile for the pulse shapes was evaluated with special attention paid to the interval of the expected $f_0$ variation after shimming (±500 Hz). Frequency response profiles were simulated using RF Pulse Wizard tool(34) ignoring relaxation. We compared pulse shapes with a small time-bandwidth product to maintain a short TE. The frequency response profile of a rectangular pulse (default for the non-selective T1-FFE sequence) and an asymmetric (single side-lobe) gauss-filtered sinc pulse were evaluated more closely. The simulations were performed for a flip angle of 16° (from experiment 1) and with identical maximum $B_1^+$ for the two pulse shapes. A map of the $f_0$ offset was acquired on an example subject. Based on this map and the frequency response profile, the reduction in flip angle was simulated.

**Experiment 3: RF pulse duration**

The influence of incidental MT effects on the $T_1$ estimation was varied via the duration of the asymmetric sinc pulse as τ=210, 700, 2000 µs. In all scans, the first echo of the readout was omitted to accommodate the longest pulse. The nominal flip angle pair was kept constant and both RF pulses used identical τ within the DFA acquisition. $T_1$ estimates were evaluated through histogram analysis and compared to literature values.

**Phantom validation**

To validate the $T_1$ estimates, 13 gel samples with unique $T_1$ values (taken from the Eurospin II set, Diagnostic Sonar Ltd) were attached to a 2000 ml flask containing Marcol™ oil, thus reducing $B_1^+$ inhomogeneities compared to in vivo. As a reference $T_1$ measurement, a single slice, IR-prepared, multi-shot EPI sequence with 3 k-space lines acquired per shot was used to measure 9 inversion times of TI=120, 200, 400, 600, 900, 1500, 2100, 3000, 4000 ms at TR=10 s. Other acquisition parameters were as follows: In-plane voxel size of 2.50×2.50 mm², slice thickness of 4.50 mm, bandwidth in the phase-encoding direction of 404 Hz/pixel, transverse FOV of 200×200 mm², SENSE-factor of 2.5, and TE=7.68 ms. Total scan time was 20 min. $T_1$ estimates were then derived by mono-exponential three-parameter fitting of the signal dependence on TI. These reference $T_1$ estimates were compared to those acquired with the DFA protocol for the separate gel samples through ROI analysis.

**In vivo validation**

The protocol was also validated in vivo. To reduce motion sensitivity, a single-shot EPI was used with a resulting echo train length of 29, bandwidth in the phase-encoding direction of 55 Hz/pixel and TE=22.83 ms. The shortest and longest TI were excluded due to incomplete MT after inversion(35) and incomplete relaxation after readout. The total scan time was 2 min 20 s. Correction of EPI distortions were done in-plane in the phase encoding direction using FSL FUGUE. After upsampling the EPI-based $T_1$ map to 0.83 x 0.83 x 0.9 mm resolution the DFA-based $T_1$ map was co-registered and then down-sampled to the original 2D EPI to mimic partial volume effects before a voxel-wise comparison.

To study the conformity of $T_1$ estimates in WM, cortical gray matter (GM) and ventricular CSF, segmentation was performed using FAST(36). To avoid partial volume effects and noise voxels, probability maps were masked so that only voxels with a probability of 1 were included and then eroded by one voxel. Guided by segmentation, manual ROIs were placed symmetrically (right-left) and compared in the caudate nucleus, putamen and thalamus.

**Cohort data**

$T_1$ maps of nine healthy volunteers (25 to 52 years old, four males and five females) were obtained with the optimized protocol. To analyze $T_1$ in various brain tissues and to assess the validity of the estimates, ROIs were manually defined in the frontal WM, frontal cortical GM, caudate nucleus, thalamus, putamen, globus pallidus and ventricular CSF. Efforts were made to place the ROIs as in the work by Rooney et al.(37).

## Results

**Experiment 1: Nominal flip angle pair and spoiling bias**

Signal bias caused by incomplete spoiling and manifested as deviations from the expected linearity of Eq. [4] was analyzed in a high and a low $f_T$ area (Figure 1). Deviations can be seen for nominal flip angles > 16° in the high $B_1^+$ area ($f_T = 1.23$) but not until 32° in the low $B_1^+$ area ($f_T = 0.73$). This corresponds to local flip angles of 16° · 1.23 ≈ 20° and 32° · 0.73 ≈ 23°, respectively.

A map of the apparent Ernst angle obtained from flip angles ≤ 16° with an accompanying whole-brain histogram is shown in Figure 2. The apparent Ernst angle is dominated by the influence of $B_1^+$ inhomogeneities, with some local contrast between GM, WM and CSF. Accordingly, the whole-brain histogram shows a smooth variation with no well-defined tissue specific modes. The whole-brain median of the apparent Ernst angle was $\alpha_E = 9.5°$ which would yield an optimal nominal flip angle pair of $\alpha_{T1}/\alpha_{PD} = 23°/4°$ regarding noise propagation. Due to the incomplete spoiling observed in high $B_1^+$ areas however, $\alpha_{T1} = 23°$ was reduced to $\alpha_{T1} = 16°$.

**Experiment 2: RF pulse shape**

Bloch equation simulations of the two evaluated RF pulse shapes can be seen in Figure 3. With the maximum $B_1^+$ amplitude of 4.317 µT, the asymmetric sinc pulse shows a flatter frequency response around the center, despite the shorter pulse duration of the rectangular pulse. For this pulse, simulations based on an $f_0$ offset map showed a reduction in local flip angle by 1° to 6° (6 to 63 % of nominal flip angle) in 6 % of the brain voxels in orbitofrontal and inferior temporal regions. Therefore, the asymmetric sinc was preferred over the rectangular pulse.

**Experiment 3: RF pulse duration**

The effect of varying the RF pulse duration, τ, on estimated $T_1$ maps can be seen in Figure 4. An increase in estimated $T_1$ when decreasing τ is visible both in the color maps and in the histograms. For example, the peak $T_1$ estimates in the WM mode were 1128, 1253, 1373 ms for τ=2000, 700, 210 µs, respectively corresponding to nominal $B_1^+$ amplitudes of 1.50, 4.32, 14.26 µT for $\alpha_{T1}$, and 0.38, 1.08, 3.57 µT for $\alpha_{PD}$. A smaller increase in the GM mode was observed while CSF appeared unaffected. Since the values obtained with the 700 µs asymmetric sinc pulses were well in agreement to literature and could later be validated using IR-prepared EPI, this pulse duration was chosen for the protocol.

**Phantom validation**

Figure 5 shows a comparison of the $T_1$ estimates in each of the 13 gel samples acquired with either the IR-prepared EPI or the DFA protocol. Across the samples, $T_1$ estimates varied between approximately 500 ms and 1700 ms, with good agreement between IR-EPI and DFA. However, a small systematic underestimation of DFA-derived $T_1$ estimates is observable with a mean deviation of -1.6 %.

**In-vivo validation**

A 2D comparison of an IR-derived and DFA-derived downsampled $T_1$ map of an example subject can be seen in Figure 6. The maps are accompanied by a linear regression plot and a Bland-Altman plot. The regression line has a slope of 0.90 and the Bland-Altman plot shows a mean deviation in $T_1$ of +0.5 % and a standard deviation of ±34 %. The ROI analysis (Table 1) shows that DFA $T_1$ values correspond well to the IR-EPI results in segmented white matter (+2 %), thalamus (+1 %) and putamen (±0 %), but were slightly lower in cortical gray matter (-3 %) and caudate nucleus (-4 %). Both methods show rather high standard deviations in CSF.

**Cohort data**

Regional $T_1$ values across nine healthy subjects are presented in Table 2, together with literature values(14,37-39). The estimates derived in this study are within the span of previously reported estimates, except for cortical GM (6 % lower than the mean of previous estimates). Figure 7 illustrates a typical 3D $T_1$ map by two sets of orthogonal views, one centered on the basal ganglia and the other on the motor cortex. A distinct distribution of $T_1$ with GM can be seen. The histogram shows clearly delineated modes representing WM (~1280 ms) and GM (~1830 ms), whereas the broad CSF mode is discernible around 4000 ms. Finally, Figure 8 shows maps of $T_1$, $R_2^*$, PD and $\chi$, to illustrate the multi-parametric capability of the finalized protocol.

# Discussion

In this study, we report on whole-brain $T_1$-mapping at 7T using the DFA approach with special focus on addressing known sources of bias by the choice of RF excitation. Firstly, the influence of incomplete spoiling was mitigated by limiting the higher flip angle and applying a post-processing correction algorithm. Secondly, the use of an RF pulse shape with a flat frequency response profile removed the influence of $B_0$ inhomogeneities. Thirdly, the duration of the RF pulse was adjusted so that magnetization transfer between the bound and free water pool was minimized. Bias minimization is particularly important when using only two flip angles, since it is impossible to identify bias from just two data points(20). Hence, in experiment 1, the flip angle was varied over a wide range. Due to the non-selective implementation, there was no need to account for slice profile effects on the $T_1$ estimates as described in a 2D VFA study at 7T(14).

$T_1$ quantification by the finalized protocol was validated: **1)** in a multi-$T_1$ phantom using a gold standard IR-prepared EPI protocol; **2)** in a single subject, again comparing to an IR-derived reference; and **3)** in a cohort of healthy volunteers, relating the $T_1$ estimates obtained by ROI analysis to the literature. In all these experiments, the DFA protocol generally yielded estimates well in agreement with the respective reference. The only notable deviation was observed in cortical GM, where the DFA protocol resulted in 3 % lower $T_1$ estimates relative the IR-prepared EPI protocol and 6 % lower than the mean of previously reported estimates(14,37-39). Regarding validation experiment 2, even though attempts were made to mimic the partial volume effects obtained in the low-resolution EPI images through down-sampling (increasing cortical GM $T_1$ estimates), it is possible that this was not compensated for fully due to slice profile effects and an imperfect co-registration. On a similar note, the single-slice studies with larger slice thickness(14,37) have reported higher cortical GM estimates than studies with smaller slice thickness(38,39) implying that partial volume effects from the CSF has led to overestimation of cortical GM $T_1$.

Although DFA-based $T_1$ quantification appears fairly straightforward to implement, it is more sensitive to $B_1^+$ inhomogeneities than the IR-based measurements often used as reference(37,39). Local flip angles typically varied between 30 % and 130 % of the nominal flip angle across the brain. Biases introduced by the more severe $B_1^+$ inhomogeneity compared to 3T are corrected post-hoc with accurate flip angle mapping(24). However, there are some exceptions in areas of low $B_1^+$ amplitude such as the temporal lobes where $T_1$ appears underestimated. The $B_1^+$ inhomogeneity could possibly be mitigated using dielectric pads(40) and/or multi-channel transmit RF technology. The choice of only two flip angles for whole brain coverage thus implies a compromise to allow shorter scan times and to make the protocol compatible with the hMRI(13) toolbox. Of course, it is possible to optimize the flip angle pair for $B_1^+$ and $T_1$ in target regions, sacrificing performance in other areas. In this work, a general optimization of the whole brain was aimed for, based on the median apparent Ernst angle.

Based on RF pulse simulations, we chose to change the RF pulse shape from a rectangular pulse to an asymmetric sinc pulse, to obtain a flat frequency response within the typical $f_0$ range in the brain at 7T. With a rectangular pulse, a sufficiently broad frequency response entailed a high $B_1^+$ amplitude (close to 20 uT) and thus considerably stronger MT effects (not shown). Contrary to the standard settings, the RF pulse duration, τ, was kept constant between the two flip angles, rather than the maximum $B_1^+$ amplitude, $B_{1,max}^+$. This approach was chosen because, for a certain pulse, it allows to balance the partial saturation of the free and the bound pool over a range of sufficiently small flip angles and thus reduce incidental MT effects(18). The partial saturation of the free water pool is $\delta_f = 1 - \cos\alpha \approx \alpha^2/2$. When the saturation of the bound pool is small, it can be approximated by $\delta_b \approx q\alpha^2/\tau$ where $q$ depends on the RF pulse shape. Thus, with $\tau$ being the same for the two pulses, $\tau$ can be chosen to match the saturation of each pool independent of flip angle, $q\alpha^2/\tau = \alpha^2/2 \Leftrightarrow q/\tau = 1/2$. When instead $B_{1,max}^+$ is kept constant, the saturation of the bound pool can be approximated as $\delta_b \approx qB_{1,max}^+\alpha$ and the balance with $\delta_f$ can only be obtained for a single flip angle, $qB_{1,max}^+\alpha = \alpha^2/2$. This will result in MT effects in the $T_1$ maps which are difficult to control, especially with large $B_1^+$ inhomogeneities.

MT effects on $T_1$ estimation have rarely been considered(41,42), but the present results show that the influence on $T_1$ estimates can be about 10%. Experiment 3 aimed to determine the $\tau$ which minimizes this MT-related bias. Too short pulses disproportionately saturated the bound pool relative the free pool, leading to MT from the free pool and thus a decrease in the steady-state signal and, in turn, an overestimation of $T_1$. Conversely, too long pulses create the opposite situation, leading to an underestimation of $T_1$. For the asymmetric sinc pulse, a duration of $\tau = 700$ μs yielded $T_1$ estimates consistent with literature and IR-prepared EPI. Still, $T_1$ was systematically underestimated in WM of the inferior temporal lobes and cerebellum, where $B_1^+$ was quite low (~50%). Since the composite $f_T$ maps were specifically designed for a wide range of $B_1^+$, the tentative explanation is that the approximation of MT given above may be an oversimplification. While our empirical adjustment of $\tau$ gives correct $T_1$ for moderate deviations of $B_1^+$, it may still imply bias where the local $B_1^+$ power is low. Despite this, overall $T_1$ estimates in WM agrees well with those obtained using IR-prepared EPI (Table 1) and literature (Table 2). Assessment of $T_1$ in different cortical regions will be covered elsewhere.

Hypothetically, the small flip angle approximation ($\alpha \ll 1$ rad) could have contributed to the deviation from linearity in experiment 1. At the highest local flip angle of 39° in Figure 1 ($\alpha_{nom} = 32°$ and $f_T = 1.23$) the deviation to the exact solution(19) is merely 4%. At 20 degrees local flip angle, the deviation to the exact solution is below 1 % which is much smaller than the observed deviation.

Apart from attempting to avoid effects from incomplete spoiling altogether by limiting the upper flip angle, we also applied a linear correction algorithm(30). Generally, the effect of this correction on the $T_1$ estimates were small compared to MT effects. Yet, in central regions of high $B_1^+$ such as the splenium and thalamus, $T_1$ estimates were reduced, and better matched the estimates observed with IR-prepared EPI and literature. Consistent with this, high $B_1^+$ areas showed deviation from the Ernst equation at a nominal flip angle of 16° (experiment 1). The correction method was developed with 3T data in mind and simulations were therefore performed on a limited range of $T_1$s from 700 to 1800 ms(30). Although barely, this should cover the range of $T_1$s found in the basal ganglia in the center of the brain where $B_1^+$ is at its highest. It would perhaps be possible to rely completely on the post-correction thus not reducing the higher flip angle. It was deemed more prudent, however, to limit the effect of residual transverse coherences at a modest cost in SNR(19). Another way to ensure sufficient spoiling could have been to exploit the long TR at 7T to implement long crusher gradients(43) instead of doing a multi-echo readout. Again, since the effect of residual transverse coherences can be effectively limited by the upper flip angle and/or correction, subject comfort and the benefit of the $R_2^*$ and susceptibility quantification may be considered more important. Longer $T_1$ times at 7T favor a longer TR compared to 3T, which was used to accommodate a multi-echo readout. This leads to improved SNR(15) with the added benefit of additional parameter quantification from the evolution of transverse magnetization (Figure 8). The timing of the echo train was similar to an established 3T multi-parametric protocol(3,8) but the shorter $T_2^*$ at 7T (~15 – 25 ms) will yield a higher precision in the $T_2^*$ estimates. The PD values obtained by the hMRI toolbox appeared reasonable despite the conversion from $|A|$ to PD being optimized for 3T. The hMRI toolbox assigns a PD of 69 p.u. to WM, which could result in errors especially in very young, elderly, or diseased subjects where a large part of WM deviates from this value(8). In combination with a MT-w acquisition, the maps of $T_1$ and $|A|$ can be used to calculate the MT saturation(44), although this is challenging at 7T due to specific absorption rate restrictions(45). Thus, accurate $T_1$ mapping will reduce the bias of $|A|$ and hence MT saturation. In addition to magnitude-based multi-parametric data, χ maps were estimated from the phase images by QSM reconstruction(32). A detailed discussion of these multi-parametric 7T maps is outside the scope of this paper, however.

Another $T_1$-mapping technique is MP2RAGE(38), which has become popular due to the inherent compensation of $B_1^+$ inhomogeneity and interleaved acquisition of the $T_1$-w and PD-w images. Derivation of $T_1$ is performed using a look-up table, and the readout train will somewhat limit the capability of simultaneous mapping of $R_2^*$ and $\chi$(46). The main benefit of the VFA approach for whole-brain $T_1$-mapping is the speed of acquisition due to the lack of recovery intervals. When compared to inversion recovery-based techniques, this reduction of scan time is particularly large at 7T due to the prolongation of $T_1$. Measurement time can thus be spent to increase spatial resolution.

## Conclusions

$T_1$-mapping at 7T using a DFA-based approach in a multi-parameter mapping context is feasible, but care should be taken to address systematic bias due to residual transverse coherences, incomplete excitation in areas with high $B_0$ deviations, $B_1^+$ inhomogeneities as well as incidental on-resonance MT effects. It is suggested to limit the upper flip angle to avoid incomplete spoiling, using an RF pulse shape that is insensitive to $B_0$ inhomogeneities as well as adjusting the RF pulse duration to reduce incidental MT effects.

## Acknowledgments

Lund Bioimaging Center (LBIC), Lund University, is acknowledged for experimental resources (equipment grant VR RFI 829-2010-5928). The project was funded by the Swedish Research Council (NT 2014-6193). Nicola Spotorno is thanked for assistance with the QSM processing.

# Tables

Table 1. **Regional $T_1$ data from a single subject (Figure 6) derived using either DFA or IR-EPI.** Mean ± SD within ROI. ROIs were automatically segmented for WM, cortical GM and CSF and manually drawn for caudate nucleus, thalamus and putamen. The high standard deviation in the CSF reflects that the flip angle pair is ill-chosen for such long $T_1$.

| Tissue | DFA $T_1$ [ms] | IR-EPI $T_1$ [ms] |
|---|---|---|
| White Matter | 1256 ± 89 | 1235 ± 82 |
| Cortical Gray Matter | 1867 ± 164 | 1928 ± 158 |
| Caudate Nucleus | 1735 ± 86 | 1810 ± 108 |
| Thalamus | 1669 ± 124 | 1647 ± 145 |
| Putamen | 1614 ± 56 | 1618 ± 82 |
| CSF | 3976 ± 681 | 3956 ± 766 |

Table 2. **Regional $T_1$ data averaged across nine subjects compared to literature.** Mean ± SD in ms across subjects. Note that Rooney et al.(37) and Wright et al.(39) used fitting of the signal dependence on TI while Marques et al.(38) used a look-up table. Overall, good agreement is observed considering the large variation in previously reported estimates.

| Tissue | This study | Rooney(37) Look-Locker | Marques(38) MP2RAGE | Wright(39) MPRAGE | Dieringer(14) 2D-VFA |
|---|---|---|---|---|---|
| White Matter | 1218±44 | 1220±36 | 1150±60 | 1130±100 | 1284±22 |
| Cortical Gray Matter | 1898±43 | 2132±103 | 1920±160 | 1939±150 | 2065±69 |
| Caudate Nucleus | 1686±64 | 1745±64 | 1630±90 | 1684±76 | - |
| Thalamus | 1659±124 | 1656±84 | 1430±100 | - | - |
| Putamen | 1646±84 | 1700±66 | 1520±90 | 1643±167 | - |
| Globus Pallidus | 1415±81 | 1347±52 | - | - | - |
| CSF | 4435±432 | 4425±137 | - | - | - |

# Figures

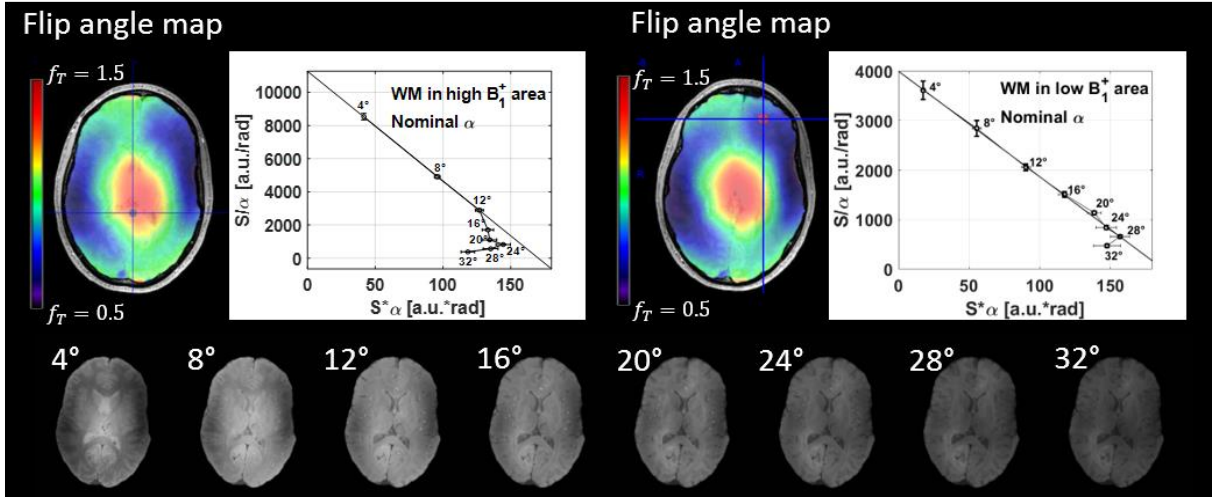

Figure 1. Linear plot of VFA signal in a high $B_1^+$ ROI ($f_T = 1.23$) situated in the splenium (left panel) and a low $B_1^+$ ROI ($f_T = 0.73$) situated in frontal WM (right panel). The line fitted to $\alpha = 4°, 8°, 12°$ is plotted (as in Eq. [4]) to highlight deviations from the Ernst equation. Residual transverse coherences become apparent at nominal flip angles of 16° and 32° in the low and high $B_1^+$ area, respectively. Taking $f_T$ into account, these nominal flip angles converts to a local flip angle of approximately 20°. The bottom row shows the spoiled gradient echo signal for each nominal flip angle in a transversal slice.

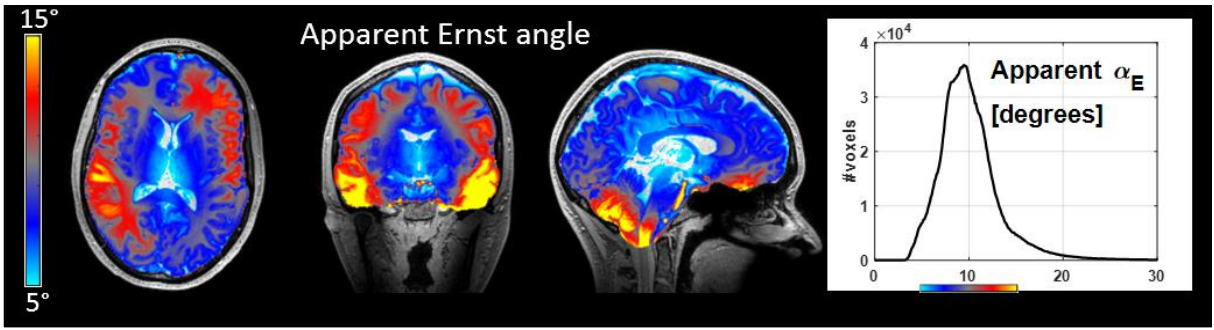

Figure 2. Apparent Ernst angle, i.e., with effects of $B_1^+$ inhomogeneity included, displayed as a pseudo-color map and a whole-brain histogram. The whole-brain median of $\alpha_E = 9.5°$ would suggest a nominal flip angle pair of $\alpha_{PD} = 4°$ and $\alpha_{T1} = 23°$ to minimize noise propagation. Note the large variations due to $B_1^+$ inhomogeneities confounding the $T_1$ dependence.

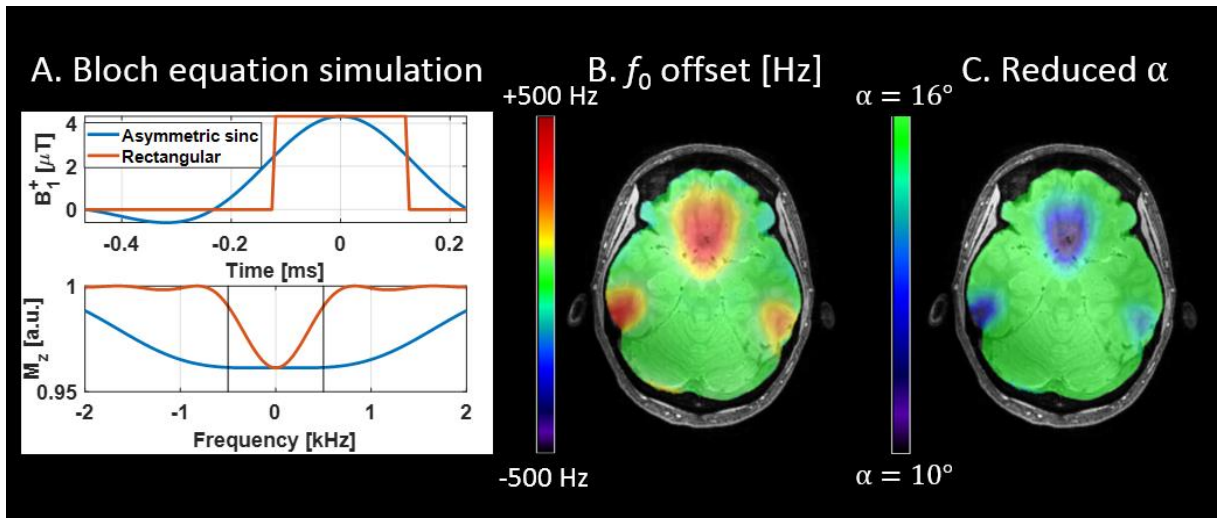

Figure 3. **A:** Bloch equation simulation of two RF pulse shapes both yielding a 16° flip angle. **Top row:** An asymmetric sinc (blue) and a rectangular pulse (red). Both pulses have a maximum amplitude of 4.317 µT but different durations of 697.6 µs and 241.8 µs respectively. **Bottom row:** Frequency response profile of the longitudinal magnetization ($M_z$) obtained with the respective pulses. The asymmetric sinc shows a homogeneous response across the expected $f_0$ offset variation across the brain at 7T (±500 Hz, indicated by black lines). Compare to the narrow response profile for the rectangular pulse. **B:** $f_0$ offset map of a subject confirming the existence of areas with $f_0$ offsets >300 Hz (red). **C:** Simulation showing the corresponding reduction in flip angle of applying the rectangular pulse. Note that these cannot be mapped by DREAM.

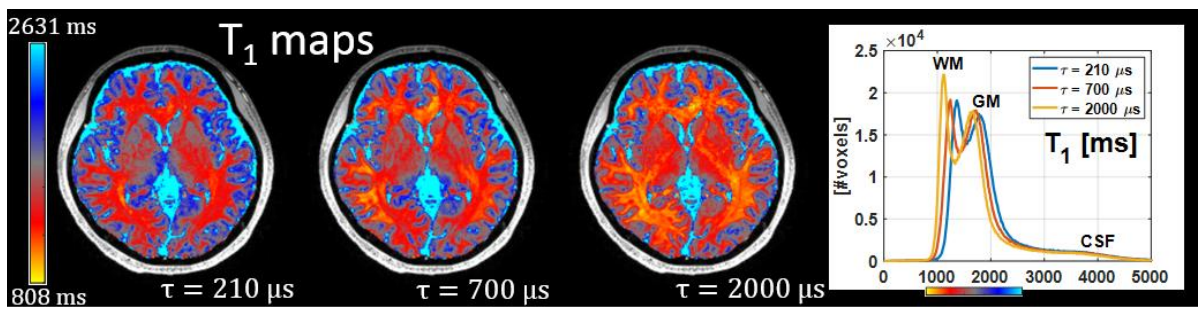

Figure 4. $T_1$ maps with accompanying whole-brain histograms showing the effect of the RF pulse duration on the $T_1$ estimation. Pulse durations were $\tau = 210\ \mu s$ (blue), $\tau = 700\ \mu s$ (red) and $\tau = 2000\ \mu s$ (yellow). Pronounced over-/underestimation of $T_1$ is evident when $\tau$ is low/high, respectively, both in the pseudo-color maps and in the WM/GM modes of the histograms. This is most likely due to incidental on-resonance MT effects caused by high/low power integral pulses. The effect is stronger in the highly myelinated WM compared to GM and nonexistent in the CSF which lacks macro-molecular content.

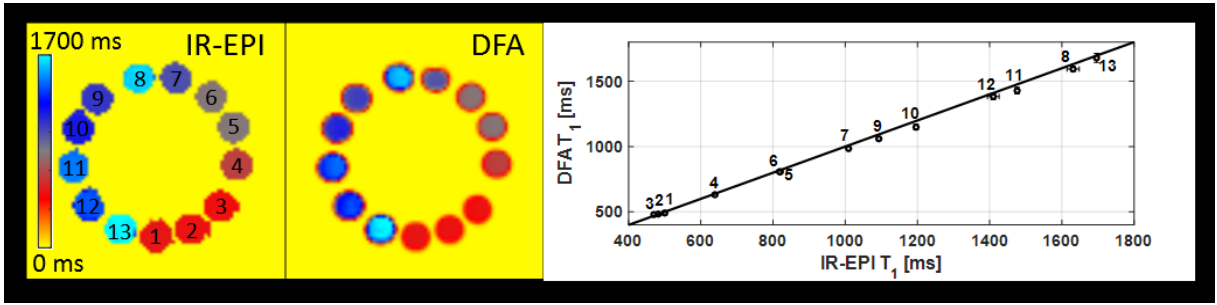

Figure 5. $T_1$ estimates in 13 gel samples, derived by either IR-prepared EPI sequences with differing TIs or the DFA protocol. The line of identity reveals a small underestimation of DFA-derived estimates compared to IR-EPI with a mean deviation of -1.6 %.

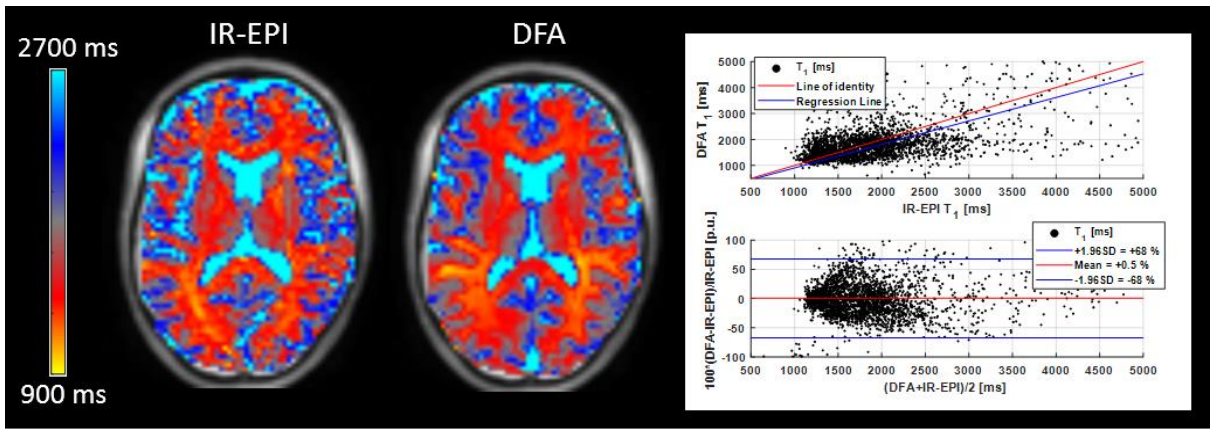

Figure 6. In vivo $T_1$ maps derived by either IR-prepared EPI sequences or the DFA protocol. The top scatter plot yielded regression line with a slope of 0.90 (upper row) and a Bland-Altman plot (lower row) with a mean deviation of DFA-derived estimates of +0.5 % and a standard deviation of ±34 %.

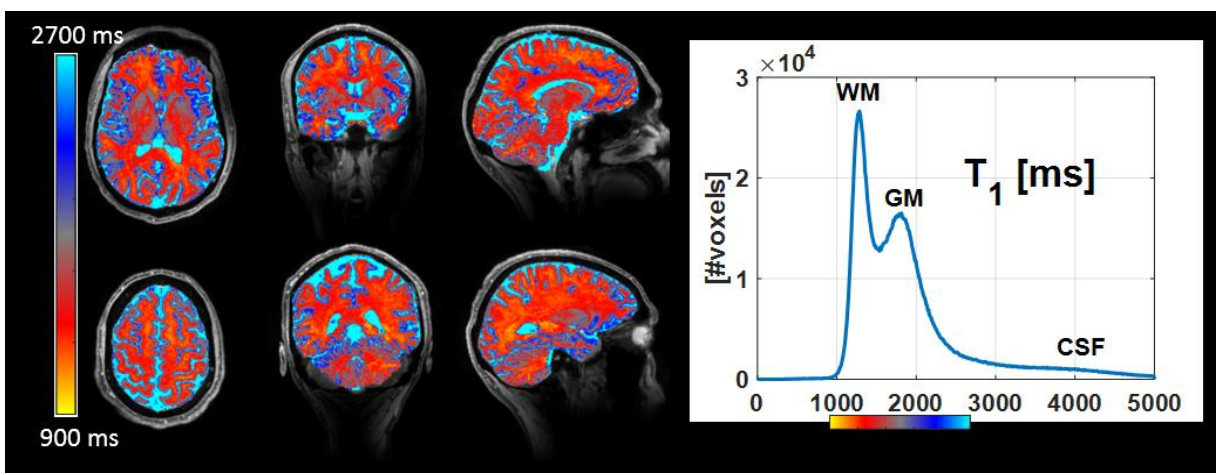

Figure 7. $T_1$ maps of a representative subject with accompanying whole-brain histogram using the finalized protocol. The color scale is centered on the GM peak (gray) to highlight variations across GM represented by gray/dark blue. WM appears red/bright orange and the CSF is shown in light blue. Orthogonal views in the upper row is centered on the basal ganglia and the lower on the motor cortex. The asymmetry in the right cerebellum is caused by the two-channel transmit/receive coil locally yielding weak $B_1^+$ coverage and low SNR. The whole brain $T_1$ histogram shows clearly delineated WM/GM modes and long tail corresponding to voxels of predominantly CSF.

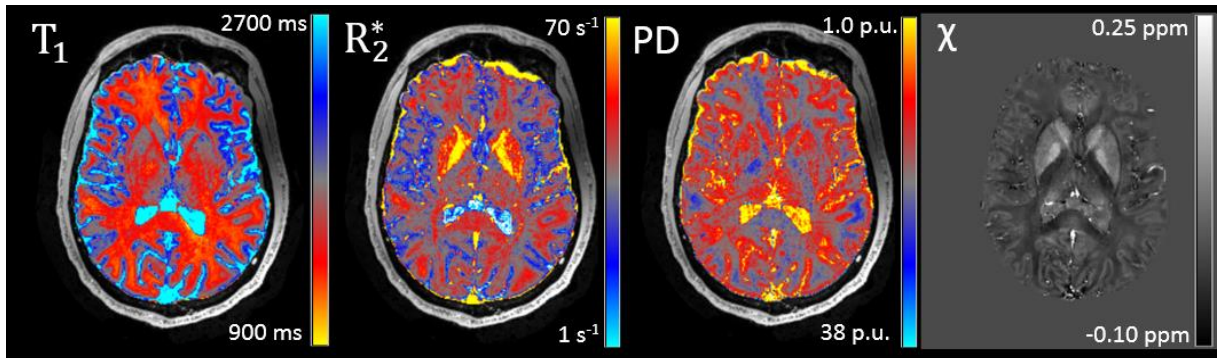

Figure 8. Example multi-parametric maps of $T_1$, $R_2^*$, PD and $\chi$ from a representative subject using the finalized protocol. Note, the well-defined structures in the visual cortex on the $\chi$ map compared to the $R_2^*$ map.